\documentclass[]{emulateapj}

\usepackage{graphicx,times}

\begin{document}

\title{Exploring the helium core of the $\delta$ Scuti star CoRoT 102749568 with asteroseismology}
\author{Xinghao Chen\altaffilmark{1,2,3},Yan Li \altaffilmark{1,2,3,4},Guifang Lin\altaffilmark{1,2}, Yanhui Chen\altaffilmark{2,5,6}, and Junjun Guo\altaffilmark{1,2,3}}

\altaffiltext{1}{Yunnan Observatories, Chinese Academy of Sciences, P.O. Box 110, Kunming 650011, China; chenxinghao@ynao.ac.cn; ly@ynao.ac.cn}
\altaffiltext{2}{Key Laboratory for Structure and Evolution of Celestial Objects, Chinese Academy of Sciences, P.O. Box 110, Kunming 650011, China}
\altaffiltext{3}{University of Chinese Academy of Sciences, Beijing 100049, China}
\altaffiltext{4}{Center for Astronomical Mega-Science, Chinese Academy of Sciences, 20A Datun Road, Chaoyang District, Beijing, 100012, China}
\altaffiltext{5}{Institute of Astrophysics, Chuxiong Normal University, Chuxiong 675000, China}
\altaffiltext{6}{School of Physics and Electronical Science, Chuxiong Normal University, Chuxiong 675000,China}

\begin{abstract}
Based on regularities in rotational splittings, we seek possible multiplets for the observed frequencies of CoRoT 102749568. Twenty-one sets of multiplets are identified, including four sets of multiplets with $l=1$, nine sets of multiplets with $l=2$, and eight sets of multiplets with $l=3$. In particular, there are three complete triplets ($f_{10}$, $f_{12}$, $f_{14}$), ($f_{31}$, $f_{34}$, $f_{35}$), and ($f_{41}$, $f_{43}$, $f_{44}$). The rotational period of CoRoT 102749568 is estimated to be $1.34^{+0.04}_{-0.05}$ days. When doing model fittings, three $l=1$ modes ($f_{12}$, $f_{34}$, and $f_{43}$) and the radial first overtone $f_{13}$ are used. Our results shows that the three nonradial modes ($f_{12}$, $f_{34}$, and $f_{43}$) are mixed modes, which mainly provide constraints on the helium core. The radial first overtone $f_{13}$ mainly provides constraint on the stellar envelope. Hence the size of the helium core of CoRoT 102749568 is determined to be $M_{\rm He}$ = 0.148 $\pm$ 0.003 $M_{\odot}$ and $R_{\rm He}$ = 0.0581 $\pm$ 0.0007 $R_{\odot}$. The fundamental parameters of CoRoT 102749568 are determined to be $M$ = 1.54 $\pm$ 0.03 $M_{\odot}$, $Z=$ 0.006, $f_{\rm ov}$ = 0.004 $\pm$ 0.002, $\log g$ = 3.696 $\pm$ 0.003, $T_{\rm eff}$ = 6886 $\pm$ 70 K, $R$ = 2.916 $\pm$ 0.039 $R_{\odot}$, and $L$ = 17.12 $\pm$ 1.13 $L_{\odot}$.
\end{abstract}


\section{Introduction}

Thanks to the space missions, MOST (Walker et al. 2003), CoRoT (Baglin et al. 2006), and $Kepler$ (Borucki et al. 2010), many $\delta$ Scuti stars are observed precisely (e.g., HD 144277 (Zwintz et al. 2011), HD 50844 (Poretti et al. 2009), and KIC 9700322 (Breger et al. 2011)). In particular, a large number of pulsation frequencies are detected in the light curves of some $\delta$ Scuti stars, such as HD 174936 (Garc{\'{\i}}a Hern{\'a}ndez et al. 2009), HD 50870 (Mantegazza et al. 2012), and HD 174966 (Garc{\'{\i}}a Hern{\'a}ndez et al. 2013). Due to the complexity of the frequency content, it is very difficult to disentangle the whole spectra of $\delta$ Scuti stars. Recently, Papar$\acute{\rm o}$ et al. (2016) developed a sequence search method, and found a large number of series of quasi-equally spaced frequencies in 77 $\delta$ Scuti stars. Besides, Chen et al. (2016) attempted to interpret the frequency spectra of the $\delta$ Scuti star HD 50844 using the rotational splitting.

CoRoT 102749568 was observed from 24 October 2007 to 3 March 2008 ($\Delta T = 131$ days) by CoRoT during the first long run in the anti-centre direction (LRa01). Guenther et al. (2012) classified the $\delta$ Scuti star CoRoT 102749568 as an F1 IV star on the basis of the low-resolution $R$ = 1300 spectra, which were observed in January 2009 with AAOmega multi-object spectrograph mounted on Anglo-Australian 3.9-m Telescope.

Papar$\acute{\rm o}$ et al. (2013) converted the spectral type F1 IV of CoRoT 102749568 into effective temperature $T_{\rm eff}$ and gravitational acceleration log$g$ using the calibrated values from Straizys $\&$ Kuriliene (1981), and then obtained $T_{\rm eff}$ $=$ $7000\pm200$ K and log$g$ $=$ $3.75\pm0.25$ by means of fitting AAOmega spectra with stellar atmosphere models of Kurucz (Kurucz 1979). Moreover, Papar$\acute{\rm o}$ et al. (2013) obtained $\upsilon\sin i$ $=$ 115 $\pm$ 20 km $\rm s^{-1}$ from the high-resolution $R$ = 85000 spectra, which were observed with Mercator Echelle Spectrograph mounted on 1.2$-$m Mercator Telescope of Roque de los Muchachos Observatory. Furthermore, Papar$\acute{\rm o}$ et al. (2013) extracted a total of 52 independent pulsation frequencies from the CoRoT timeseries. These frequencies are listed in Table 1. They identified the oscillation frequency 9.702 $\rm d^{-1}$ with the largest amplitude as the radial first overtone with the method of multi-colour photometry. Moreover, Papar$\acute{\rm o}$ et al. (2013) identified other 11 frequencies based on the regularities in frequency spacings.

Mode identification is very important for the asteroseismic study of pulsation stars. For a rotating star, the regularities due to rotational splitting in observed frequencies help us much to identify their spherical harmonic degree $l$ and azimuthal number $m$. Based on the rotational splitting law of g modes, we successfully disentangled the frequency spectra of the $\delta$ Scuti star HD 50844 (Chen et al. 2016). That motivates us to analyze another $\delta$ Scuti star CoRoT 102749568 with the same method. In Section 2, we propose our mode identification by means of rotational splitting. In Section 3, we describe the details of input physics and model calculations. Input physics are described in Section 3.1, model grids are elaborated in Section 3.2, and the optimal model are analyzed in Section 3.3. We discuss our results in Section 4, and summarize the results in Section 5.

\section{Mode identification based on rotational splitting}
A pulsation mode is characterized by three indices, i.e., the radial order $k$, the spherical harmonic degree $l$, and the azimuthal number $m$ (Christensen-Dalsgaard 2003). The azimuthal number $m$ are degenerate for a spherically symmetric star. Namely, modes with the same $k$ and $l$ but different $m$ have the same frequency. Stellar rotation will break the structure of spherical symmetry and result in frequency splitting, i.e., one nonradial pulsation frequency will spilt into $2l+1$ different frequencies. According to the theory of stellar oscillation, a general formula for rotational splitting is described as (Aerts et al. 2010)
\begin{equation}
\nu_{k,l,m}= \nu_{k,l} + \beta_{k,l}\frac{m}{P_{\rm rot}}.
\end{equation}
In Equation (1), $\beta_{k,l}$ is the rotational parameter measuring the size of rotational splitting and $P_{\rm rot}$ the rotational period. For high-degree or high-order p modes, $\beta_{k,l}\simeq$ l. Values of rotational splitting for pulsation modes with different spherical harmonic degree $l$ are the same. For high-order g modes, $\beta_{k,l}\simeq$ $1-\frac{1}{l(l+1)}$ (Brickhill 1975). The rotational splitting derived from $l=1$ modes and those from $l=2$ modes and $l=3$ modes conform to the relation $\delta\nu_{k, l=1} : \delta\nu_{k, l=2}$ : $\delta\nu_{k, l=3}$ = 0.6 : 1: 1.1 (Winget et al. 1991). Based on these regularities in rotational splittings, we analyze the frequency spectra of CoRoT 102749568 and list possible multiplets in Table 2.

It can be noticed in Table 2 that we find twenty-one sets of multiplets, including three different types of rotational spliting. The averaged frequency splitting $\delta\nu_{1}$ is 4.451 $\mu$Hz for Multiplet 1, 2, 3, and 4. The averaged frequency splitting $\delta\nu_{2}$ is 7.453 $\mu$Hz for Multiplet 5, 6, 7, 8, 9, 10, 11, 12, and 13, and the averaged frequency splitting $\delta\nu_{3}$ is 8.176 $\mu$Hz for Multiplet 14, 15, 16, 17, 18, 19, 20, and 21. For these frequency differences in Table 2, we find that some of them approximate to the corresponding averaged value $\delta\nu_{1}$, $\delta\nu_{2}$, or $\delta\nu_{3}$ (e.g., Miltiplet 1, 2, 3, and 5), and some of them are several times that of the corresponding average value (e.g., Multiplet 4, 11, 12, and 13). Moreover, we find that the ratio of $\delta\nu_{1}$ : $\delta\nu_{2}$ : $\delta\nu_{3}$ is 0.597 : 1.0 : 1.097, which agrees well with the property of g modes. As shown in Figure 1, the $\delta$ Scuti star CoRoT 102749568 is in the post-main-sequence evolution stage with a contracting helium core and an expanding envelope. Such stellar structure may reproduce these behaviors of rotational splitting.

Based on the property of rotational splitting for g modes, we identify frequencies in Multiplet 1, 2, 3, and 4 as $l=1$ modes, frequencies in Multiplet 5, 6, 7, 8, 9, 10, 11, 12, and 13 as $l=2$ modes, and frequencies in Multiplet 14, 15, 16, 17, 18, 19, 20, and 21 as $l=3$ modes. Furthermore, we find that the azimuthal number $m$ of pulsation modes in Multiplet 1, 2, 3, 4, 6, and 13 can be uniquely identified, and the azimuthal number $m$ of pulsation modes in other multiplets allow of several possibilities (e.g., three possibilities for pulsation modes in Multiplet 5).

Finally, there are three unidentified frequencies $f_{4}$, $f_{40}$, and $f_{52}$, which do not show frequency splitting. Frequencies $f_{4}$ and $f_{10}$ have a difference of 29.789 $\mu$Hz, about four times that of $\delta\nu_{k,l=2}$. However, $f_{10}$ has been regarded as one component of Multiplet 1. Multiplet 1 consists of three components, being a complete triplet. Frequencies $f_{10}$ and $f_{12}$ have a difference of 4.485 $\mu$Hz, which agree well with the difference 4.399 $\mu$Hz between $f_{12}$ and $f_{14}$. Besides, modes with lower degree $l$ are easier to be observed because of the effect of geometrical cancellation. Frequencies $f_{27}$ and $f_{40}$ have a difference of 48.847 $\mu$Hz, about six times that of $\delta\nu_{k,l=3}$. Similarly, the frequency $f_{27}$ has been identified as one component of Multiplet 14. Frequencies $f_{48}$ and $f_{51}$ have a difference 29.712 $\mu$Hz, about four times that of $\delta\nu_{k,l=2}$. Frequencies $f_{48}$ and $f_{52}$ have a difference of 40.597 $\mu$Hz, about five times that of $\delta\nu_{k,l=3}$. The spherical harmonic degree of $f_{48}$ allows of two possibilities, $l=2$ or $l=3$. For the former case, the azimuthal number $m$ of $f_{48}$ and $f_{51}$ are determined to be $m=$ $(-2, +2)$. This case is listed in Table 2. The azimuthal number of the latter case allows of two possibilities, i.e., $m=$ $(-3, +2)$ and $(-2, +3)$.

Based on the above analyses, the detection of triplets, quintuplets, and septuplets helps us to identify four sets of multiplets with $l=1$, nine sets of multiplets with $l=2$, and eight sets of multiplets with $l=3$. Owing to the deviations from the asymptotic expression, we find in Table 2 that slight differences of the rotational splitting exist in different multiplets (e.g., in Multiplet 8 and 9). Besides, slight differences also exist in the same multiplet (e.g., in Multiplet 1). Furthermore, we find in Table 2 that there are only two components in Multiplet 4, 7, 8, 9, 10, 11, 12, 13, 15, 16, 17, 18, 19, 20, and 21. Different physical origins are also possible, such as the large separation led by the so-called island modes (Garc{\'{\i}}a Hern{\'a}ndez et al. 2013; Ligni{\`e}res et al. 2006) and the phenomenon of avoided crossings (Aizenman et al. 1977). The $\delta$ Scuti star CoRoT 102749568 is a slightly evolved star, the occurrence of avoided crossings will make the frequency spectra more complex. From the observed frequency spectra of CoRoT 102749568, it is difficult to find the signs of avoided crossings. Hence the phenomenon of avoided crossings is not considered in our work.

\section{Input physics and model calculations}
\subsection{Input physics}
All of our theoretical models are computed with the Modules for Experiments in Stellar Astrophysics (MESA), which is developed by Paxton et al. (2011, 2013). We use the so-called module pulse from version 6596 to calculate stellar evolutionary models and their corresponding pulsation frequencies (Christensen-Dalsgaard 2008; Paxton et al. 2011, 2013)

In our calculations, the OPAL opacity table GS98 (Grevesse $\&$ Sauval 1998) series are adopted. We use the $T - \tau$ relation of Eddington grey atmosphere in the atmosphere integration, and choose the mixing-length theory (MLT) of B$\ddot{\rm o}$hm-Vitense (1958) to treat convection. Based on numerical calculations, we find that theoretical evolutionary models are not sensitive to the mixing-length parameter. However, the values of $\beta_{k,l}$ of theoretical models with slightly higher $\alpha_{\rm MLT}$ agree better with asymptotic values of g modes, hence $\alpha_{\rm MLT}$ = 2.2 is adopted in our work. Moreover, we find that theoretical models without convective core overshooting can not reproduce those observed multiplets. Hence we introduce the convective core overshooting in our calculations.. For the overshooting mixing of the convective core, we adopt an exponentially decaying prescription. Following Freytag et al. (1996) and Herwig (2000), we introduce an overshoot mixing diffusion coefficient
\begin{equation}
D_{\rm ov}=D_{0}exp(\frac{-2 z}{f_{\rm ov} H_{\rm p}}).
\end{equation}
In Equation (2), $D_{0}$ is the convective mixing coefficient, $z$ the distance into radiative zone away from the boundary of convective core, $H_{\rm p}$ the pressure scale height, and $f_{\rm ov}$ an adjustable parameter describing the efficiency of the overshooting mixing. In our calculations, we set the lower limit of the diffusion coefficient $D_{\rm ov}^{\rm limit}=1\times10^{-2}$ $\rm cm^{2}/\rm s$, below which no element mixing is allowed. In addition, effects of rotation and element diffusion are not considered in our work.
\subsection{Model grids}
The internal structure and the evolutionary track of a star depend on the initial mass $M$, the initial chemical composition $(X,Y,Z)$, and the overshooting parameters $f_{\rm ov}$. A grid of theoretical models are computed with MESA, $M$ varying from 1.5 $M_{\odot}$ to 2.2 $M_{\odot}$ with a step of 0.01 $M_{\odot}$, $Z$ varying from 0.005 to 0.030 with a step of 0.001, and $f_{\rm ov}$ varying from 0 to 0.016 with a step of 0.001. In our calculations, we choose the initial helium fraction $Y=0.245+1.54Z$ (e.g., Dotter et al. 2008; Thompson et al. 2014; Tian et al. 2015) as a function of mass fraction of heavy-elements $Z$.

Theoretical models for each star are computed from the zero-age main sequence to post-main-sequence stage. The error box in Figure 1 corresponds to the observed stellar parameters, i.e., the effective temperature 6800 K $<$ $T_{\rm eff}$ $<$ 7200 K and the gravitational acceleration 3.50 $<$ log$g$ $<$ 4.00. We calculate frequencies of oscillation modes with $l$ = 0, 1, 2, and 3 for every stellar model which falls inside the error box along the evolutionary track.

\subsection{Optimal models}
We try to use theoretical oscillation frequencies derived from a grid of evolutionary models to fit those of identified pulsation modes. According to the analyses in Section 2, mode identifications are unique only in Multiplet 1, 2, and 3 and their $m=0$ components are observed. When doing model fittings, we hence use four identified pulsation modes, i.e., three $l=1$ modes ($f_{12}$, $f_{34}$, and $f_{43}$) and the radial first overtone $f_{13}$. Papar$\acute{\rm o}$ et al. (2013) identify the frequency $f_{13}$ with the largest amplitude as the radial first overtone with the method of multi-colour photometry. In our calculations, we use the identification of $f_{13}$ as the radial first overtone. When doing model fittings, we use the following criterion
\begin{equation}
\chi^{2}=\frac{1}{n}\sum(|\nu_{i}^{\rm obs}-\nu_{i}^{\rm theo}|^{2}),
\end{equation}
where $\nu_{i}^{\rm obs}$ is the observed frequency, $\nu_{i}^{\rm theo}$ the theoretically calculated frequency, and $n$ the amount of the observed frequencies.

Figure 2 shows the plot of $1/\chi^{2}$ to the effective temperature $T_{\rm eff}$ for all theoretical models. Each curve in Figure 2 corresponds to one theoretical evolutionary track. In Figure 2, the filled circles correspond to seven candidate models of CoRoT 102749568 in Table 3.

Christensen-Dalsgaard (2003) defines the general expression of the rotational parameter $\beta_{k,l}$ of a pulsation mode for a rigid body as
\begin{equation}
\beta_{k,l}=\frac{\int_{0}^{R}(\xi_{r}^{2}+L^{2}\xi_{h}^{2}-2\xi_{r}\xi_{h}-\xi_{\rm h}^{2})r^{2}\rho dr}
{\int_{0}^{R}(\xi_{r}^{2}+L^{2}\xi_{h}^{2})r^{2}\rho dr},
\end{equation}
where the subscripts "$r$" and "$h$" correspond to the radial displacement and the horizontal displacement, $\rho$ denotes the local density, and $L^{2}= l(l+1)$. Based on the asymptotic behavior of the eigenfunctions of high-order g modes, $\beta_{k,l}$ can be simplified as being the asymptotic value $1-\frac{1}{L^{2}}$. According to asymptotic value of g modes, $\beta_{k,l=1}$ = 0.5, $\beta_{k,l=2}$ = 0.833, and $\beta_{k,l=3}$ = 0.917.

For the three identified $l=1$ modes $f_{12}$, $f_{34}$, and $f_{43}$, their corresponding $\beta_{k,l}$ of these candidate modes are listed in Table 4. In Table 4, rotational parameters $\beta_{k,l}$ of observed frequencies are asymptotic values of g modes based on Equation (1). It can be found in Table 4 that theoretical values of $\beta_{k,l}$ for Model 1, 2, 3, and 7 significantly deviate from those asymptotic values of g modes. We therefore exclude these four models from our considerations. The physical parameters of CoRoT 102749568 are obtained based on Model 4, 5, and 6. These parameters are listed in Table 5. In our work, we select the theoretical model (Model 4) with minimum value of $\chi^{2}$ = 0.016 as the optimal model. Its theoretical evolutionary track corresponds to the curve in Figure 1.

Theoretical pulsation frequencies of the optimal model are listed in Table 6, in which $n_{\rm p}$ is the amount of radial nodes in the propagation of p modes and $n_{\rm g}$ the amount of radial nodes in the propagation of g modes. We notice in Table 6 that most of the pulsation modes are gravity and mixed modes. Figure 3 shows the plot of $\beta_{k,l}$ to theoretical pulsation frequencies for the optimal model. We can find in Figure 3 that most of $\beta_{k,l}$ are in good agreement with the asymptotic value of g modes. These pulsation modes possess more pronounced g-mode features. Besides, there are several pulsation modes, whose $\beta_{k,l}$ obviously deviate from the asymptotic values of g modes. They have more pronounced p-mode features.

Comparisons of results of pulsation frequencies in Table 2 are listed in Table 7. The $m\ne 0$ pulsation frequencies in columns denoted with $\nu^{\rm theo}$ are derived from $m=0$ modes according to Equation (1). The filled circles in Figure 3 correspond to $m=0$ components of the multiplets in Table 7. It can be noticed in Table 7 that $m=0$ components in Multiplet 1, 2, 3, 5, 9, 10, 14, and 18 are observed, while $m=0$ components in Multiplet 4, 6, 7, 8, 11, 12, 13, 15, 16, 17, 19, 20, and 21 are absent. In Figure 3, we notice that $\beta_{k,l}$ of $m=0$ components in Multiplet 1, 2, 3, 5, 9, 10, 14, and 18 agree well with the asymptotic value of g modes. For Multiplet 4, 6, 7, 8, 11, 12, 15, 16, 17, 19, 20, and 21, $\beta_{k,l}$ of their corresponding $m=0$ components are also in accordance with Equation (1). Moreover, we find that $\beta_{k,l}$ of corresponding $m=0$ components in Multiplet 13 are slightly larger than the asymptotic value from Equation (1). These results also proves that our approach of mode identification based on the rotational splitting of g modes is self-consistent.

Finally, we try to do mode identification for the three isolated pulsation frequencies based on the optimal model, and list them in Table 8. We notice in Table 8 that there are two possible model counterparts for $f_{4}$, i.e., $(1,0,-51,-1)$ 76.394 $\mu$Hz and $(2,0,-117,+2)$ 76.303 $\mu$Hz. For $f_{40}$, $(2,3,-35,+2)$ 190.803 $\mu$Hz may be possible model counterpart. According the analyses in Section 2, the spherical harmonic degree of $f_{48}$ allows of two possibilities, i.e., $l=2$ and $l=3$. When identifying $f_{48}$ and $f_{51}$ as being two  $l=2$ modes, their possible model counterparts are listed in Table 7. If $f_{48}$ and $f_{52}$ are identified as being two $l=3$ modes, there are no suitable model counterparts for them.
\section{Discussions}
When doing model fittings, we use four identified pulsation modes including the radial first overtone $f_{13}$ and three $l=1$ modes ($f_{12}$, $f_{34}$, and $f_{43}$). Figure 4 shows the propagation diagram of the optimal model. Based on the default parameters, we adopt the position where the hydrogen fraction $X_{cb}$ = 0.01 as the boundary of the helium core. The outer zone is the stellar envelope, and the inner zone is the helium core. The vertical curves in Figures 4 and 5 indicate the boundary of the helium core. Figure 5 shows the scaled eigenfunctions of the radial first overtone and the three $l=1$ nonradial pulsation modes. It can be seen clearly in Figure 5 that the radial first overtone mainly propagates in the stellar envelope, and therefore mainly provide constraints on the stellar envelope. For the three nonradial pulsation modes, Figure 5 shows that they behave g-mode features in the helium core and p-mode features in the stellar envelope. Then the three nonradial pulsation modes mainly provide constraints on the helium core.

Following Chen et al. (2016), we introduce two asteroseismic parameters, i.e., the acoustic radius $\tau_{0}$ and the period separation $\Pi_{0}$. The acoustic radius $\tau_{0}$ is a significant physical parameter in the asteroseismic study. The acoustic radius $\tau_{0}$ carries information on the stellar envelope (e.g., Ballot et al. 2004; Miglio et al. 2010; Chen et al. 2016). The acoustic radius $\tau_{0}$ is defined as (Aerts et al. 2010)
\begin{equation}
\tau_{0}=\int_{0}^{R}\frac{dr}{c_{\rm s}},
\end{equation}
in which $R$ is the stellar radius and $c_{\rm s}$ the adiabatic sound speed. According to Equation (5), the value of acoustic radius $\tau_{0}$ is mainly dominated by the profile of $c_{\rm s}$ inside the stellar envelope.

According to the theory of stellar oscillations, g-mode oscillations are gravity waves. They mainly propagate inside the helium core. Their properties can be characterized by $\Pi_{0}$, which is defined as
\begin{equation}
\Pi_{0}= 2\pi^{2}(\int_{0}^{R}\frac{N}{r}dr)^{-1}
\end{equation}
(Unno et al. 1979; Tassoul 1980; Aerts et al. 2010), where $N$ is the Brunt-V$\ddot{\rm a}$is$\ddot{\rm a}$l$\ddot{\rm a}$ frequency. According to Equation (6), $\Pi_{0}$ is mainly dominated by the profile of Brunt-V$\ddot{\rm a}$is$\ddot{\rm a}$l$\ddot{\rm a}$ frequency $N$ inside the helium core.

To fit the four pulsation modes ($f_{12}$, $f_{13}$, $f_{34}$, and $f_{43}$), both the helium core and the stellar envelope of the theoretical model need to be matched to the actual structure of CoRoT 102749568. It can be found in Table 3 that $\tau_{0}$ and $\Pi_{0}$ of the three preferred models (Model 4, 5, and 6) are very close. This is because they are nearly alike in structure. Thus the size of the helium core of CoRoT 102749568 is determined to be $M_{\rm He}$ = 0.148 $\pm$ 0.003 $M_{\odot}$ and $R_{\rm He}$ = 0.0581 $\pm$ 0.0007 $R_{\odot}$. The errors are estimated on basis of the deviations of the helium cores of Model 5 and 6 from that of Model 4.

According to Equation (1), the rotational period $P_{\rm rot}$ of the $\delta$ Scuti star CoRoT 102749568 is determined to be $P_{\rm rot}$ = $1.34^{+0.04}_{-0.05}$ days. Meanwhile, we find in Table 5 that the theoretical radius $R$ of CoRoT 102749568 is 2.916 $\pm$ 0.039 $R_{\odot}$. According to $\upsilon_{\rm rot} = 2\pi R/P_{\rm rot}$, the rotational velocity at the equator is then deduced to be $\upsilon_{\rm rot}$ = $109.8^{+6.4}_{-4.6}$ km $\rm s^{-1}$, which is in agreement with the value of $\upsilon \sin i$ = 115 $\pm$ 20 km $\rm s^{-1}$ (Papar$\acute{\rm o}$ et al. 2013).

It should be noticed that Equation (1) only contains the first-order effect of rotation $C_{1}m/P_{\rm rot}$, in which $C_{1}$ = $1-\frac{1}{L^{2}}$. The second-order effect of rotation is derived by Dziembowski $\&$ Goode (1992) as being $\frac{m^{2}C_{2}}{P_{\rm rot}^{2}\nu_{k,l,0}}$, the coefficient $C_{2}$ = $\frac{4L^{2}(2L^{2}-3)-9}{2L^{4}(4L^{2}-3)}$.  The ratio of the second-order effect and the first-order effect is then deduced to be $\phi_{l}= \frac{C_{2}}{C_{1}} \frac{m}{P_{\rm rot}\nu_{k,l,0}}$. Assuming $\nu_{k,l,0}$ = 100 $\mu$Hz, the absolute value of $\phi_{l=1}$ is estimated to be 0.0043, $\phi_{l=2}$ to be 0.0141$|m|$, and $\phi_{l=3}$ to be 0.0073$|m|$. For pulsation modes with $l=1$, the second-order effect is 0.43$\%$ that of the first-order. For pulsation modes with $l=2$ and $l=3$, the ratios $\phi_{l=2}$ and $\phi_{l=3}$ are in direct proportion to the azimuthal number $m$. The second-order effect is 2.82$\%$ that of the first-order for modes with $l=2$ and $|m|=2$. The second-order effect is $2.19\%$ that of the first-order for modes with $l=3$ and $|m|=3$. In brief, the second-order effect of stellar rotation is much less than that of the first-order. Hence the second-order effect of rotation is not considered in our work.

Finally, our model fitting results show that a slight increase in the convective core size is essential to explain these multiplets. There are two different ways to increase the convective core size, i.e., convective core overshooting (Herwig 2000; Li $\&$ Yang 2007; Zhang 2013) and rotation (Eggenberger et al. 2010; Girardi et al. 2011). Maeder $\&$ Meynet (2000) and Yang et al. (2013) found that the effects of rotation on stellar structure and evolution depend on the masses of stellar models. Moreover,Yang et al. (2013) noticed that 2.05 $M_{\odot}$ is a critical mass. Rotation results in an increase in the convective core size for stars with $M$ $>$ 2.05 $M_{\odot}$. The effect is similar to that of the convective core overshooting. However for stars with $M$ $<$ 2.05 $M_{\odot}$, rotation leads to a decrease in the convective core size. The optimal model in our work corresponds to a star with $M=1.54M_{\odot}$, $Z=0.006$, $f_{\rm ov} = 0.004$. According to the analyses of Yang et al. (2013), rotation will result in a slight decrease in the convective core size. If the effects of rotation are included in theoretical evolutionary models, a larger convective core overshooting may be indispensable.
\section{Summary}
In this work, we carry out asteroseismic analyses and numerical calculations for the $\delta$ Scuti star CoRoT 102749568. The main results are concluded as follows:

1. We identify twenty-one sets of multiplets using the regularities in rotational splittings, including four sets of multiplets with $l=1$, nine sets of multiplets with $l=2$, and eight sets of multiplets with $l=3$. In particular, there are three complete triplets, i.e., ($f_{10}$, $f_{12}$, $f_{14}$), ($f_{31}$, $f_{34}$, $f_{35}$), and ($f_{41}$, $f_{43}$, $f_{44}$). The rotational period $P_{\rm rot}$ is estimated to be $1.34^{+0.04}_{-0.05}$ days according to the frequency differences in these multiplets.

2. Based on our model calculations, the $\delta$ Scuti star CoRoT 102749568 is in post-main-sequence evolution stage. The stellar parameters of the $\delta$ Scuti star CoRoT 102749568 are determined to be $M$ = 1.54 $\pm$ 0.03 $M_{\odot}$, $Z=$ 0.006, $f_{\rm ov}$ = 0.004 $\pm$ 0.002, $\log g$ = 3.696 $\pm$ 0.003, $T_{\rm eff}$ = 6886 $\pm$ 70 K, $R$ = 2.916 $\pm$ 0.039 $R_{\odot}$, and $L$ = 17.12 $\pm$ 1.13 $L_{\odot}$.

3. Based on our optimal model, we notice that most of the oscillation frequencies are mixed modes. The radial first overtone $f_{13}$ mainly provides constraints on the stellar envelope. The three nonradial pulsation modes $f_{12}$, $f_{34}$, and $f_{43}$ possess more pronounced g-mode features, which mainly provide constraints on the helium core. The property of the stellar envelope is characterized by the acoustic radius $\tau_{0}$, and the property of the helium core is characterized by the period separation $\Pi_{0}$. Finally, the size of the helium core of CoRoT 102749568 is determined to be $M_{\rm He}$ = 0.148 $\pm$ 0.003 $M_{\odot}$ and $R_{\rm He}$ = 0.0581 $\pm$ 0.0007 $R_{\odot}$.

\acknowledgments
This work is funded by the NSFC of China (Grant No. 11333006, 11521303, 11503079, and 11563001) and by the foundation of Chinese Academy of Sciences (Grant No. XDB09010202). The authors gratefully acknowledge the computing time granted by the Yunnan Observatories, and provided on the facilities at the Yunnan Observatories Supercomputing Platform. The authors are sincerely grateful to an anonymous referee for instructive advice and productive suggestions. The authors are also very grateful to the suggestions from Q.-S. Zhang, T. Wu, and J. Su.

  \begin{figure}
  \epsscale{1.0}
  \plotone{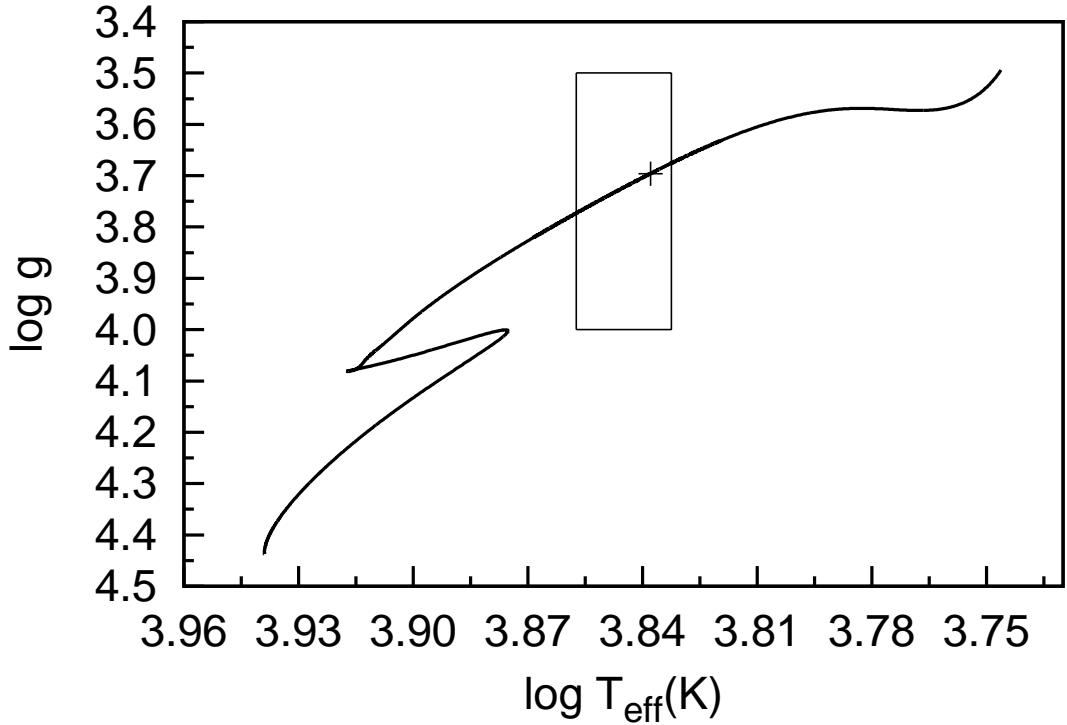}
  \caption{Evolutionary track of $M$ =1.54 $M_{\odot}$, $Z$=0.006, $f_{\rm ov}$=0.004. The rectangle corresponds to the error box of the observed parameters, 3.5 $<$ $\log g$ $<$ 4.0 and 6800 K $<$ $T_{\rm eff}$ $<$ 7200 K. The cross marks the location of the optimal model (Model 4).}
  \label{Figure.1}
  \end{figure}

  \begin{figure}
  \epsscale{1}
  \plotone{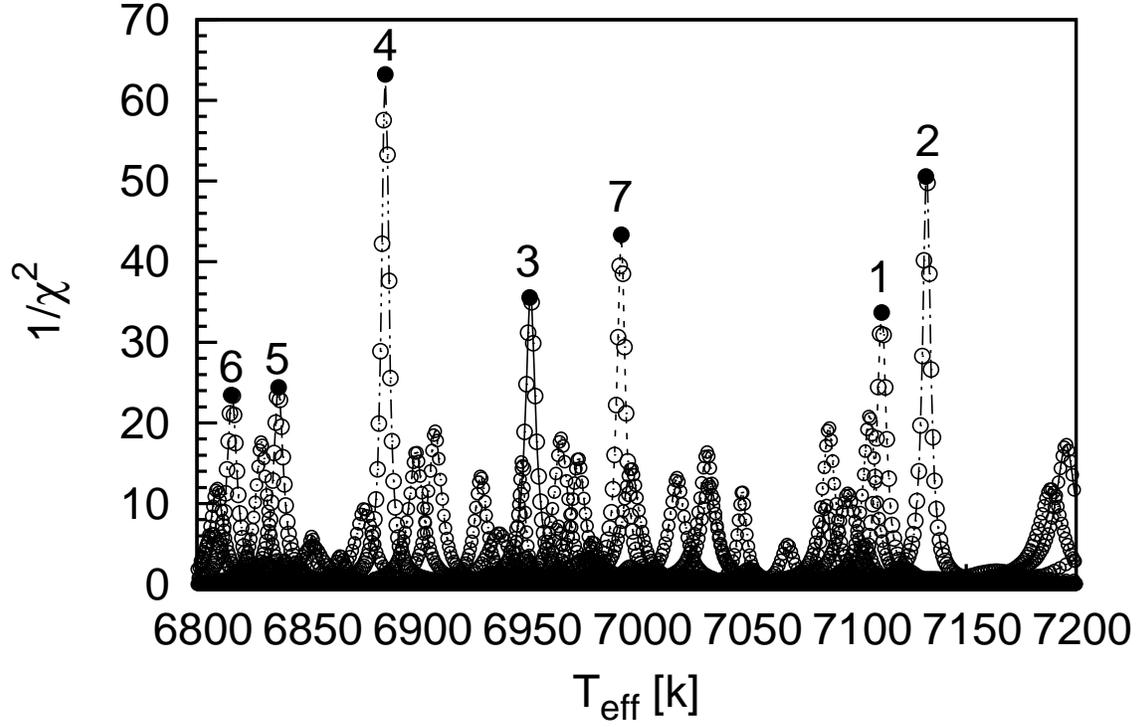}
  \caption{Plot of $1/\chi^{2}$ to the effective temperature $T_{\rm eff}$ of all theoretical models. The filled circles indicate the candidate models in Table 3.}
  \label{Figure.2}
  \end{figure}

  \begin{figure}
  \epsscale{1}
  \plotone{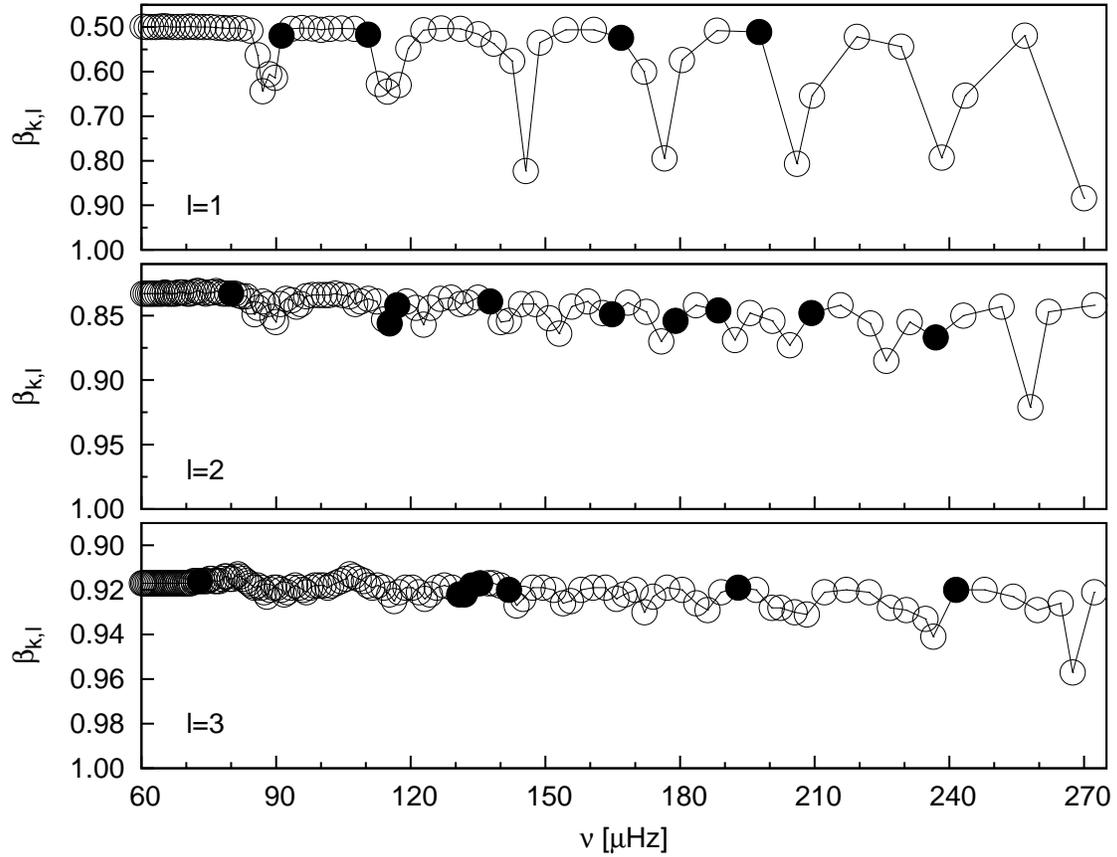}
  \caption{Plot of $\beta_{k,l}$ versus theoretically calculated frequency $\nu$ of the optimal model. The filled circles correspond to $m=0$ components of the multiplets in Table 7.}
  \label{Figure.3}
  \end{figure}

  \begin{figure}
  \epsscale{1}
  \plotone{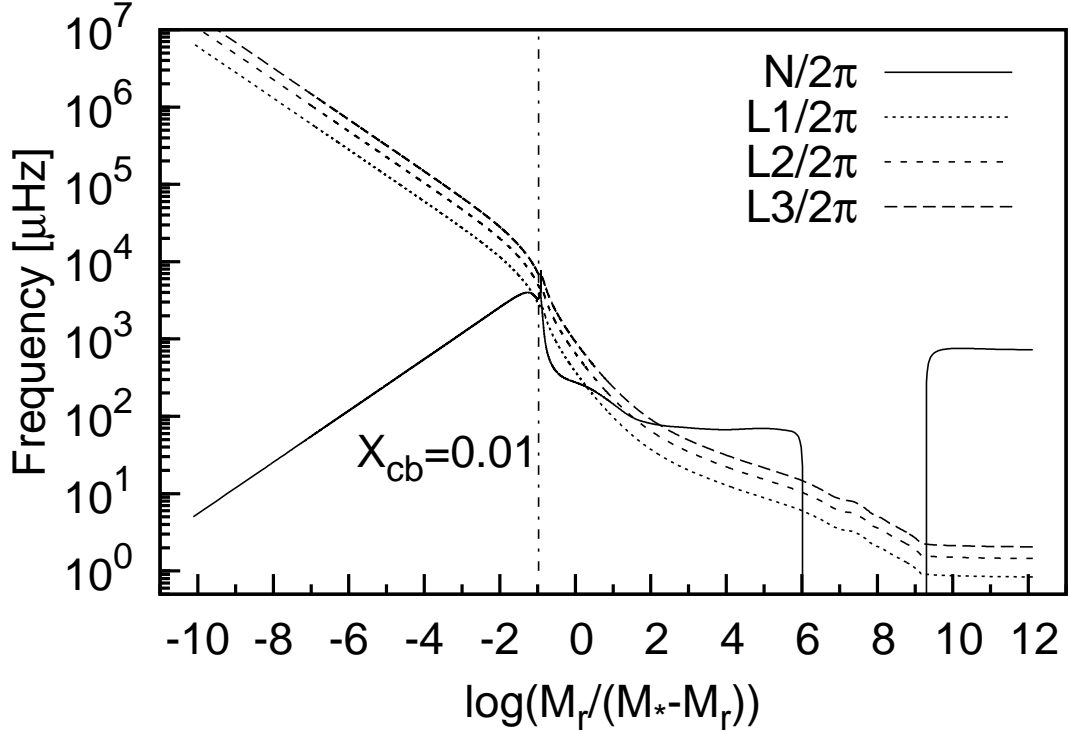}
  \caption{$N$ is Brunt$-$V$\ddot{\rm a}$is$\ddot{\rm a}$l$\ddot{\rm a}$ frequency and $L_{l}$ ($l= 1, 2, 3$) are Lamb frequency. $M_{*}$ is the stellar mass. The vertical line indicates the boundary of the helium core.}
  \label{Figure.4}
  \end{figure}

  \begin{figure}
  \epsscale{1}
  \plotone{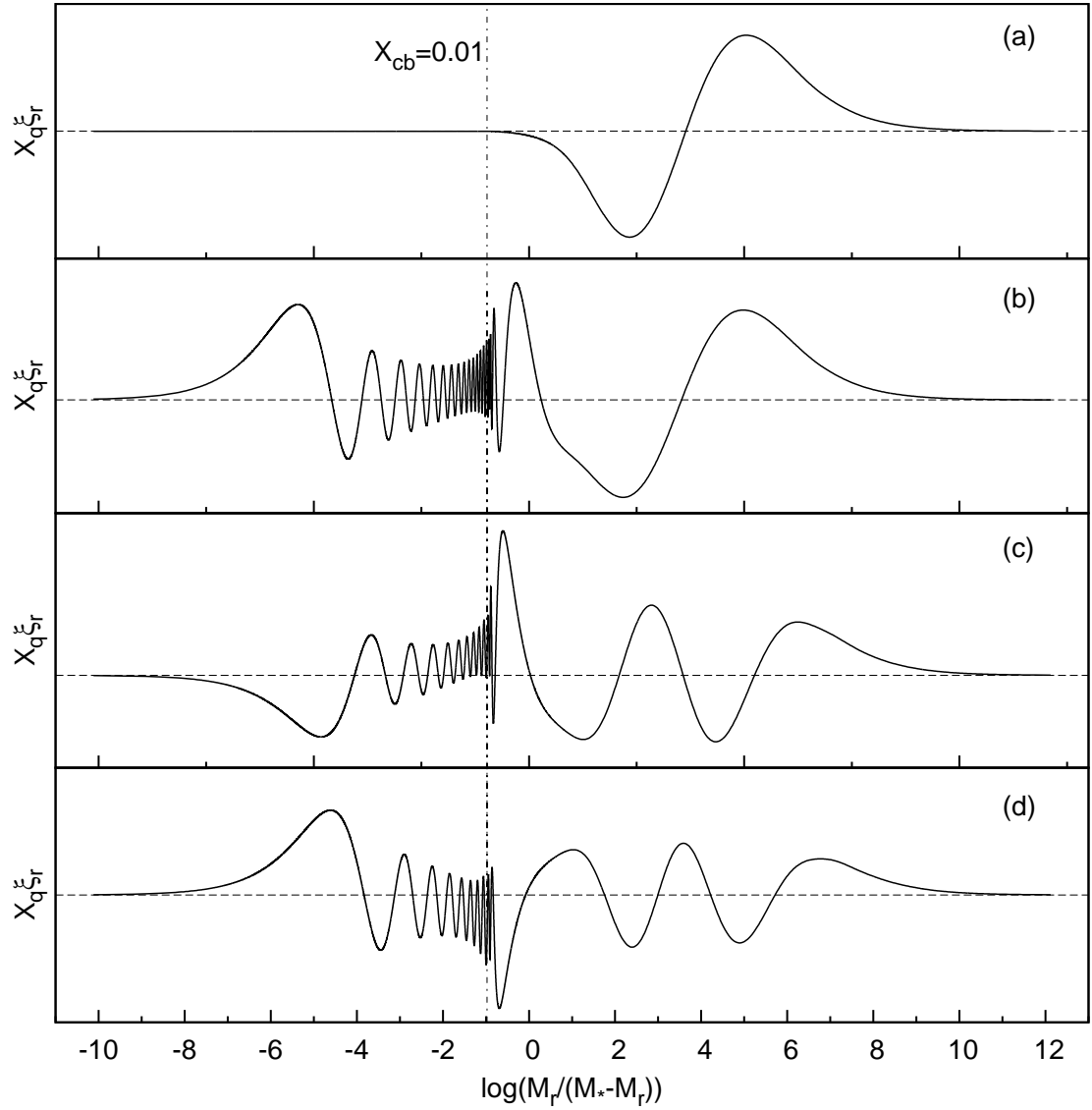}
  \caption{Scaled eigenfunctions of the radial first overtone $f_{13}$ and the three nonradial modes $f_{12}$, $f_{34}$, and $f_{43}$. $X_{q} = \sqrt{q(1-q)}$ and $q=M_{r}/M_{*}$. Panel (a) is for the radial first overtone 112.212 $\mu$Hz $(l=0,n_{p}=1,n_{g}=0)$. Panel (b) is for the mode 110.476 $\mu$Hz $(l=1,n_{p}=1,n_{g}=-37)$. Panel (c) is for the mode 166.873 $\mu$Hz $(l=1,n_{p}=3,n_{g}=-24)$. Panel (d) is for the mode 197.617 $\mu$Hz $(l=1,n_{p}=4,n_{g}=-20)$. Vertical line indicates the boundary of the helium core.}
  \label{Figure.5}
  \end{figure}

\begin{table*}
\caption{\label{t1}The 52 independent frequencies of CoRoT 102749568 obtained by Papar$\acute{\rm o}$ et al. (2013). The columns named by ID are the serial number of observed frequencies. Freq. indicates the observed frequency in unit of $\mu$Hz. Ampl. indicates the amplitude in unit of mmag.}
\centering
\begin{tabular}{lccccc}
\hline\hline
ID         &Freq.       &Ampl.    &ID         &Freq.       &Ampl.  \\
           &($\mu$Hz)   &(mmag)   &           &($\mu$Hz)   &(mmag) \\
  \hline
$f_{1}$    &64.936      &0.16     &$f_{27}$   &141.765     &1.01   \\
$f_{2}$    &65.541      &0.18     &$f_{28}$   &144.934     &0.27   \\
$f_{3}$    &72.978      &0.18     &$f_{29}$   &155.380     &0.23   \\
$f_{4}$    &76.363      &0.25     &$f_{30}$   &158.977     &0.27   \\
$f_{5}$    &87.275      &0.25     &$f_{31}$   &162.625     &0.29   \\
$f_{6}$    &96.149      &1.75     &$f_{32}$   &164.262     &0.17   \\
$f_{7}$    &96.938      &0.75     &$f_{33}$   &164.855     &0.73   \\
$f_{8}$    &100.779     &0.26     &$f_{34}$   &167.007     &0.14   \\
$f_{9}$    &102.072     &0.14     &$f_{35}$   &171.485     &0.23   \\
$f_{10}$   &106.152     &0.39     &$f_{36}$   &171.638     &0.23   \\
$f_{11}$   &108.372     &0.43     &$f_{37}$   &172.243     &0.22   \\
$f_{12}$   &110.637     &1.44     &$f_{38}$   &176.285     &0.15   \\
$f_{13}$   &112.291     &10.51    &$f_{39}$   &189.056     &0.16   \\
$f_{14}$   &115.036     &4.77     &$f_{40}$   &190.612     &0.14   \\
$f_{15}$   &115.706     &0.41     &$f_{41}$   &192.909     &0.16   \\
$f_{16}$   &115.872     &0.22     &$f_{42}$   &194.179     &0.68   \\
$f_{17}$   &117.666     &0.23     &$f_{43}$   &197.503     &0.15   \\
$f_{18}$   &122.559     &0.27     &$f_{44}$   &201.898     &0.20   \\
$f_{19}$   &122.769     &0.88     &$f_{45}$   &203.652     &0.60   \\
$f_{20}$   &123.812     &0.17     &$f_{46}$   &209.708     &0.14   \\
$f_{21}$   &124.571     &0.19     &$f_{47}$   &216.758     &0.17   \\
$f_{22}$   &125.296     &0.92     &$f_{48}$   &222.367     &0.18   \\
$f_{23}$   &132.028     &0.19     &$f_{49}$   &233.083     &0.24   \\
$f_{24}$   &133.453     &1.21     &$f_{50}$   &249.725     &0.15   \\
$f_{25}$   &134.071     &0.19     &$f_{51}$   &252.079     &0.14   \\
$f_{26}$   &134.762     &3.66     &$f_{52}$   &262.964     &0.16   \\
\hline
\end{tabular}
\end{table*}

\begin{table*}
\footnotesize
\caption{\label{t2}Possible multiplets due to stellar rotation. $\delta\nu$ = frequency difference in $\mu$Hz}
\centering
\begin{tabular}{lccccccccccc}
\hline\hline
Multiplet &ID    &Freq.   &$\delta\nu$&$l$  &$m$            &Multiplet &ID      &Freq.     &$\delta\nu$ &$l$ &$m$\\
        &        &($\mu$Hz)&($\mu$Hz) &     &               &          &        &($\mu$Hz) &($\mu$Hz)&     &   \\
\hline
        &$f_{10}$&106.152   &         &1    &$-1$           &      &$f_{18}$&122.559   &         &2    &$(-2,-1)$\\
        &        &          &4.485    &     &               &11    &        &          &22.375   &     &\\
1       &$f_{12}$&110.637   &         &1    &$0$            &      &$f_{28}$&144.934   &         &2    &$(+1,+2)$\\
        &        &          &4.399    &     &               &      &\\
        &$f_{14}$&115.036   &         &1    &$+1$           &      &\\
        &        &          &         &     &               &      &$f_{42}$ &194.179  &         &2   &$(-2,-1)$\\
        &        &          &         &     &               &12    &        &          &22.579\\
        &$f_{31}$&162.625   &         &1    &$-1$           &      &$f_{47}$ &216.758  &         &2   &$(+1,+2)$\\
        &        &          &4.382    &     &               &      &\\
2       &$f_{34}$&167.007   &         &1    &$0$            &      &\\
        &        &          &4.478    &     &               &      &$f_{48}$&222.367   &          &2   &$-2$\\
        &$f_{35}$&171.485   &         &1    &$+1$           &13    &        &          &29.712    &    &\\
        &        &          &         &     &               &      &$f_{51}$&252.079   &          &2   &$+2$\\
        &        &          &         &     &               &      &\\
        &$f_{41}$&192.909   &         &1    &$-1$           &      &\\
        &        &          &4.594    &     &               &      &$f_{22}$&125.296   &         &3   &$(-3,-2,-1,0,+1)$\\
3       &$f_{43}$&197.503   &         &1    &$0$            &      &        &          &8.157    &    &\\
        &        &          &4.395    &     &               &14    &$f_{24}$&133.453   &         &3   &$(-2,-1,0,+1,+2)$\\
        &$f_{44}$&201.898   &         &1    &$+1$           &      &        &          &8.312    &    &\\
        &        &          &         &     &               &      &$f_{27}$&141.765   &         &3   &$(-1,0,+1,+2,+3)$\\
        &        &          &         &     &               &      &\\
        &$f_{5}$ &87.275    &         &1    &$-1$           &      &\\
4       &        &          &8.874    &     &               &      &$f_{15}$&115.706   &         &3   &$(-3,-2,-1,0,+1,+2)$\\
        &$f_{6}$ &96.149    &         &1    &$+1$           &15    &       &           &8.106    &    &\\
        &        &          &         &     &               &      &$f_{20}$&123.812   &         &3  &$(-2,-1,0,+1,+2,+3)$\\
        &        &          &         &     &               &      &\\
        &$f_{8}$ &100.779   &         &2    &$(-2,-1,0)$    &      &\\
        &        &          &7.593    &     &               &      &$f_{17}$&117.666  &          &3   &$(-3,-2,-1,0,+1)$\\
5       &$f_{11}$&108.372   &         &2    &$(-1,0,+1)$    &16    &        &         &16.405    &    &\\
        &        &          &7.500    &     &               &      &$f_{25}$&134.071  &          &3   &$(-1,0,+1,+2,+3)$ \\
        &$f_{16}$&115.872   &         &2    &$(0,+1,+2)$    &      &\\
        &        &          &         &     &               &      &\\
        &        &          &         &     &               &      &$f_{49}$&233.083  &          &3   &$(-3,-2,-1,0,+1)$\\
        &$f_{9}$ &102.072   &         &2    &$-2$           &17    &        &         &16.642    &    & \\
        &        &          &22.499   &     &               &      &$f_{50}$&249.725  &          &3    &$(-1,0,+1,+2,+3)$ \\
6       &$f_{21}$&124.571   &         &2    &$+1$           &      &\\
        &        &          &7.457    &     &               &      &\\
        &$f_{23}$&132.028   &         &2    &$+2$           &      &$f_{26}$&134.762  &          &3   &$(-3,-2,-1,0)$\\
        &        &          &         &     &               &18    &        &         &24.215    &    &\\
        &        &          &         &     &               &      &$f_{30}$&158.977  &          &3   &$(0,+1,+2,+3)$\\
        &$f_{2}$ &65.541    &         &2    &$(-2,-1,0,+1)$ &      &\\
7        &       &          &7.437    &     &               &      &\\
        &$f_{3}$ &72.978    &         &2    &$(-1,0,+1,+2)$ &      &$f_{1}$ &64.936   &          &3   &$(-3,-2,-1)$\\
        &        &          &         &     &               &19    &        &         &32.002    &    &\\
        &        &          &         &     &               &      &$f_{7}$ &96.938   &          &3   &$(+1,+2,+3)$\\
        &$f_{32}$&164.262   &         &2    &$(-2,-1,0,+1)$ &      &\\
8        &       &          &7.376    &     &               &      &\\
        &$f_{36}$&171.638   &         &2    &$(-1,0,+1,+2)$ &      &$f_{19}$&122.769  &          &3   &$(-3,-2,-1)$\\
        &        &          &         &     &               &20    &        &         &32.611    &    &\\
        &        &          &         &     &               &      &$f_{29}$&155.380  &          &3   &$(+1,+2,+3)$\\
        &$f_{33}$&164.855   &         &2    &$(-2,-1,0,+1)$ &      &\\
9        &       &          &7.388    &     &               &      &\\
        &$f_{37}$&172.243   &         &2    &$(-1,0,+1,+2)$ &      &$f_{38}$&176.285  &          &3   &$(-3,-2,-1)$\\
        &        &          &         &     &               &21    &       &          &33.423    &    &\\
        &        &          &         &     &               &      &$f_{46}$&209.708  &          &3   &$(+1,+2,+3)$\\
        &$f_{39}$&189.056   &         &2    &$(-2,-1,0)$    &      &\\
10      &        &          &14.596   &     &               &      &\\
        &$f_{45}$&203.652   &         &2    &$(0,+1,+2)$    &      &\\

\hline
\end{tabular}
\end{table*}

\begin{table*}
\caption{\label{t3}Candidate models with $\chi^{2}$ $<$ 0.05 of four observed frequencies for the $\delta$ Scuti star CoRoT 102749568.}
\centering
\begin{tabular}{cccccccccccc}
\hline\hline
Model   &$Z$  &$M$          &$f_{\rm ov}$ &$T_{\rm eff}$ &log$g$ &$R$            &$L$        &$\tau_{0}$ &$\Pi_{0}$ &$\chi^{2}$\\
        &     &[$M_{\odot}$]&             &[K]           &[dex]  &$[R_{\odot}]$  &$[L_{\odot}]$  &$[\rm h]$   &$[\rm s]$ &     \\
\hline
1       &0.010&1.74         &0            &7111          &3.706  &3.065          &21.52       &4.21       &431.1    &0.030\\
2       &0.010&1.75         &0            &7132          &3.704  &3.080          &21.99       &4.23       &430.7    &0.020\\
3       &0.006&1.57         &0.001        &6951          &3.693  &2.953          &18.24       &4.14       &330.0    &0.028\\
4       &0.006&1.54         &0.004        &6886          &3.696  &2.916          &17.12       &4.08       &331.8    &0.016\\
5       &0.006&1.52         &0.005        &6837          &3.698  &2.889          &16.34       &4.03       &331.5    &0.041\\
6       &0.006&1.51         &0.006        &6816          &3.699  &2.877          &15.99       &4.01       &331.8    &0.043\\
7       &0.007&1.59         &0.013        &6993          &3.695  &2.965          &18.84       &4.16       &427.7    &0.023\\
\hline
\end{tabular}
\end{table*}

\begin{table*}
\caption{\label{t4}Theoretical rotational parameters of the three nonradial pulsation modes ($f_{12}$, $f_{34}$, and $f_{43}$) for the candidate models in Table 3. The rotational parameters $\beta_{k,l}$ of observed frequencies inside brackets are the asymptotic value of g modes according to Equation (1).}
\centering
\begin{tabular}{llllllll}
\hline\hline
Model&$f_{12}$($\beta_{k,l}$) &$f_{34}$($\beta_{k,l}$) &$f_{43}$($\beta_{k,l}$)\\
     &($\mu$Hz)               &($\mu$Hz)               &($\mu$Hz)\\
\hline
 obs &110.637(0.5)    &167.007(0.5)      &197.503(0.5)\\
 1   &110.618(0.519)  &166.977(0.528)    &197.704(0.541)\\
 2   &110.622(0.528)  &166.843(0.535)    &197.606(0.553)\\
 3   &110.411(0.538)  &167.118(0.514)    &197.421(0.537)\\
 4   &110.476(0.517)  &166.873(0.524)    &197.617(0.511)\\
 5   &110.332(0.511)  &167.215(0.513)    &197.401(0.508)\\
 6   &110.440(0.523)  &166.798(0.516)    &197.773(0.507)\\
 7   &110.753(0.573)  &167.170(0.570)    &197.499(0.522)\\
\hline
\end{tabular}
\end{table*}

\begin{table*}
\caption{\label{t5} Fundamental parameters of the $\delta$ Scuti star CoRoT 102749568.}
\centering
\begin{tabular}{lcc}
\hline\hline
Parameter               &Values \\
\hline
$M/M_{\odot}$           &1.54  $\pm$ 0.03\\
$Z$                     &0.006\\
$f_{\rm ov}$            &0.004 $\pm$ 0.002\\
$T_{\rm eff}$ $[K]$     &6886  $\pm$ 70 \\
log$g$                  &3.696 $\pm$ 0.003\\
$R/R_{\odot}$           &2.916 $\pm$ 0.039\\
$L/L_{\odot}$           &17.12 $\pm$ 1.13\\
$M_{\rm He}/M_{\odot}$  &0.148  $\pm$ 0.003\\
$R_{\rm He}/R_{\odot}$  &0.0581 $\pm$ 0.0007\\
\hline
\end{tabular}
\end{table*}

\begin{table*}
\footnotesize
\caption{\label{t6}Theoretically calculated frequencies of the optimal model. $\nu^{\rm theo}$ denotes calculated frequency in $\mu$Hz. $n_{p}$ is the amount of radial nodes in propagation cavity of p modes. $n_{g}$ is the amount of radial nodes in propagation cavity of g modes. $\beta_{k,l}$ is one rotational parameter measuring the size of rotational splitting.}
\centering
\begin{tabular}{lcccccccccc}
\hline\hline
$\nu^{\rm theo}(l,n_{p},n_{g})$ &$\beta_{k,l}$  &$\nu^{\rm theo}(l,n_{p},n_{g})$ &$\beta_{k,l}$  &$\nu^{\rm theo}(l,n_{p},n_{g})$  &$\beta_{k,l}$ &$\nu^{\rm theo}(l,n_{p},n_{g})$ &$\beta_{k,l}$ & $\nu^{\rm theo}(l,n_{p},n_{g})$ &$\beta_{k,l}$\\
($\mu$Hz)          &     &($\mu$Hz)          &     &($\mu$Hz)          &        &($\mu$Hz)       &     &($\mu$Hz)          &\\
\hline
  86.446(0, 0,   0)&     &  60.956(2, 0,-119)&0.833& 121.149(2, 1, -58)&0.844&  68.292(3, 0,-150)&0.917& 112.635(3, 1, -89)&0.918\\
 112.212(0, 1,   0)&     &  61.482(2, 0,-118)&0.833& 122.819(2, 2, -58)&0.857&  68.772(3, 0,-149)&0.917& 113.962(3, 1, -88)&0.919\\
 140.555(0, 2,   0)&     &  61.955(2, 0,-117)&0.833& 124.380(2, 2, -57)&0.844&  69.238(3, 0,-148)&0.917& 115.230(3, 1, -87)&0.922\\
 170.701(0, 3,   0)&     &  62.414(2, 0,-116)&0.833& 126.576(2, 2, -56)&0.837&  69.665(3, 0,-147)&0.917& 116.273(3, 1, -86)&0.925\\
 201.509(0, 4,   0)&     &  62.948(2, 0,-115)&0.833& 128.974(2, 2, -55)&0.836&  70.089(3, 0,-146)&0.917& 117.415(3, 1, -85)&0.921\\
 232.369(0, 5,   0)&     &  63.532(2, 0,-114)&0.833& 131.076(2, 2, -54)&0.841&  70.565(3, 0,-145)&0.917& 118.825(3, 1, -84)&0.919\\
 263.224(0, 6,   0)&     &  64.130(2, 0,-113)&0.833& 132.641(2, 2, -53)&0.840&  71.076(3, 0,-144)&0.917& 120.332(3, 1, -83)&0.919\\
                   &     &  64.703(2, 0,-112)&0.833& 134.982(2, 2, -52)&0.836&  71.596(3, 0,-143)&0.916& 121.821(3, 1, -82)&0.921\\
  70.483(1, 0, -59)&0.499&  65.208(2, 0,-111)&0.832& 137.679(2, 2, -51)&0.839&  72.092(3, 0,-142)&0.916& 123.096(3, 1, -81)&0.924\\
  71.418(1, 0, -58)&0.499&  65.726(2, 0,-110)&0.833& 140.046(2, 2, -50)&0.855&  72.543(3, 0,-141)&0.916& 124.285(3, 1, -80)&0.922\\
  72.506(1, 0, -57)&0.500&  66.333(2, 0,-109)&0.833& 141.883(2, 2, -49)&0.854&  73.009(3, 0,-140)&0.916& 125.794(3, 1, -79)&0.919\\
  73.862(1, 0, -56)&0.500&  66.987(2, 0,-108)&0.833& 144.601(2, 2, -48)&0.841&  73.536(3, 0,-139)&0.916& 127.481(3, 2, -79)&0.918\\
  75.327(1, 0, -55)&0.500&  67.649(2, 0,-107)&0.833& 147.760(2, 2, -47)&0.841&  74.095(3, 0,-138)&0.916& 129.203(3, 2, -78)&0.919\\
  76.814(1, 0, -54)&0.501&  68.269(2, 0,-106)&0.832& 150.828(2, 2, -46)&0.852&  74.655(3, 0,-137)&0.916& 130.767(3, 2, -77)&0.922\\
  78.139(1, 0, -53)&0.502&  68.812(2, 0,-105)&0.832& 153.094(2, 2, -45)&0.864&  75.177(3, 0,-136)&0.915& 132.033(3, 2, -76)&0.922\\
  79.258(1, 0, -52)&0.503&  69.408(2, 0,-104)&0.832& 155.838(2, 3, -45)&0.843&  75.656(3, 0,-135)&0.915& 133.602(3, 2, -75)&0.918\\
  80.717(1, 0, -51)&0.502&  70.101(2, 0,-103)&0.833& 159.389(2, 3, -44)&0.839&  76.178(3, 0,-134)&0.916& 135.456(3, 2, -74)&0.917\\
  82.436(1, 0, -50)&0.503&  70.836(2, 0,-102)&0.833& 162.691(2, 3, -43)&0.848&  76.763(3, 0,-133)&0.916& 137.243(3, 2, -73)&0.917\\
  84.244(1, 0, -49)&0.508&  71.568(2, 0,-101)&0.832& 164.831(2, 3, -42)&0.849&  77.372(3, 0,-132)&0.916& 138.354(3, 2, -72)&0.917\\
  85.900(1, 0, -48)&0.563&  72.226(2, 0,-100)&0.831& 168.348(2, 3, -41)&0.840&  77.966(3, 0,-131)&0.915& 139.917(3, 2, -71)&0.918\\
  86.984(1, 0, -47)&0.644&  72.820(2, 0, -99)&0.831& 172.456(2, 3, -40)&0.847&  78.498(3, 0,-130)&0.914& 141.887(3, 2, -70)&0.920\\
  88.472(1, 0, -46)&0.606&  73.523(2, 0, -98)&0.832& 175.818(2, 3, -39)&0.870&  79.017(3, 0,-129)&0.914& 143.568(3, 2, -69)&0.927\\
  89.761(1, 1, -46)&0.614&  74.320(2, 0, -97)&0.833& 178.990(2, 3, -38)&0.854&  79.615(3, 0,-128)&0.915& 145.066(3, 2, -68)&0.924\\
  91.237(1, 1, -45)&0.519&  75.147(2, 0, -96)&0.832& 183.554(2, 3, -37)&0.842&  80.264(3, 0,-127)&0.915& 147.132(3, 2, -67)&0.919\\
  93.371(1, 1, -44)&0.504&  75.942(2, 0, -95)&0.832& 188.534(2, 4, -37)&0.846&  80.910(3, 0,-126)&0.914& 149.459(3, 2, -66)&0.919\\
  95.705(1, 1, -43)&0.502&  76.619(2, 0, -94)&0.831& 192.235(2, 4, -36)&0.869&  81.486(3, 0,-125)&0.913& 151.823(3, 2, -65)&0.920\\
  98.039(1, 1, -42)&0.503&  77.309(2, 0, -93)&0.832& 195.553(2, 4, -35)&0.848&  82.016(3, 0,-124)&0.914& 153.906(3, 2, -64)&0.926\\
 100.008(1, 1, -41)&0.506&  78.153(2, 0, -92)&0.833& 200.574(2, 4, -34)&0.854&  82.643(3, 0,-123)&0.915& 155.655(3, 2, -63)&0.925\\
 101.955(1, 1, -40)&0.504&  79.068(2, 0, -91)&0.833& 204.466(2, 4, -33)&0.873&  83.332(3, 0,-122)&0.916& 157.949(3, 2, -62)&0.920\\
 104.540(1, 1, -39)&0.503&  79.981(2, 0, -90)&0.833& 209.244(2, 4, -32)&0.848&  84.004(3, 0,-121)&0.917& 160.619(3, 3, -62)&0.919\\
 107.462(1, 1, -38)&0.504&  80.774(2, 0, -89)&0.833& 215.729(2, 4, -31)&0.842&  84.576(3, 0,-120)&0.918& 163.381(3, 3, -61)&0.919\\
 110.476(1, 1, -37)&0.517&  81.481(2, 0, -88)&0.834& 222.360(2, 5, -31)&0.856&  85.165(3, 0,-119)&0.919& 165.825(3, 3, -60)&0.924\\
 112.907(1, 1, -36)&0.628&  82.374(2, 0, -87)&0.835& 225.926(2, 5, -30)&0.885&  85.875(3, 0,-118)&0.918& 167.604(3, 3, -59)&0.922\\
 114.805(1, 1, -35)&0.645&  83.381(2, 0, -86)&0.835& 231.158(2, 5, -29)&0.855&  86.630(3, 0,-117)&0.919& 170.052(3, 3, -58)&0.920\\
 117.293(1, 2, -35)&0.630&  84.384(2, 0, -85)&0.839& 236.938(2, 5, -28)&0.867&  87.352(3, 0,-116)&0.921& 172.081(3, 3, -57)&0.930\\
 119.439(1, 2, -34)&0.548&  85.209(2, 0, -84)&0.849& 243.129(2, 5, -27)&0.850&  87.978(3, 0,-115)&0.923& 173.973(3, 3, -56)&0.923\\
 122.770(1, 2, -33)&0.507&  85.979(2, 0, -83)&0.844& 251.631(2, 6, -27)&0.843&  88.642(3, 0,-114)&0.920& 177.089(3, 3, -55)&0.919\\
 126.783(1, 2, -32)&0.503&  87.005(2, 0, -82)&0.839& 258.065(2, 6, -26)&0.921&  89.426(3, 0,-113)&0.919& 180.462(3, 3, -54)&0.920\\
 131.038(1, 2, -31)&0.504&  88.122(2, 0, -81)&0.841& 262.090(2, 6, -25)&0.847&  90.259(3, 0,-112)&0.919& 183.642(3, 3, -53)&0.926\\
 135.062(1, 2, -30)&0.516&  89.156(2, 0, -80)&0.851& 272.306(2, 6, -24)&0.842&  91.077(3, 0,-111)&0.920& 186.137(3, 3, -52)&0.929\\
 138.470(1, 2, -29)&0.537&  89.961(2, 0, -79)&0.855&                   &     &  91.802(3, 0,-110)&0.922& 189.156(3, 3, -51)&0.921\\
 142.601(1, 2, -28)&0.576&  90.967(2, 0, -78)&0.841&  60.086(3, 0,-171)&0.917&  92.496(3, 0,-109)&0.921& 192.937(3, 3, -50)&0.919\\
 145.615(1, 2, -27)&0.823&  92.189(2, 0, -77)&0.837&  60.402(3, 0,-170)&0.917&  93.320(3, 0,-108)&0.919& 196.957(3, 4, -50)&0.920\\
 148.732(1, 3, -27)&0.535&  93.447(2, 1, -77)&0.838&  60.736(3, 0,-169)&0.917&  94.228(3, 0,-107)&0.918& 200.446(3, 4, -49)&0.928\\
 154.508(1, 3, -26)&0.506&  94.558(2, 1, -76)&0.843&  61.104(3, 0,-168)&0.917&  95.152(3, 0,-106)&0.919& 202.351(3, 4, -48)&0.928\\
 160.808(1, 3, -25)&0.506&  95.478(2, 1, -75)&0.841&  61.488(3, 0,-167)&0.917&  96.017(3, 1,-106)&0.920& 205.527(3, 4, -47)&0.930\\
 166.873(1, 3, -24)&0.524&  96.685(2, 1, -74)&0.835&  61.870(3, 0,-166)&0.917&  96.769(3, 1,-105)&0.921& 208.233(3, 4, -46)&0.931\\
 171.996(1, 3, -23)&0.600&  98.085(2, 1, -73)&0.834&  62.227(3, 0,-165)&0.917&  97.597(3, 1,-104)&0.919& 212.184(3, 4, -45)&0.921\\
 176.529(1, 3, -22)&0.795&  99.512(2, 1, -72)&0.834&  62.565(3, 0,-164)&0.917&  98.564(3, 1,-103)&0.918& 217.025(3, 4, -44)&0.920\\
 180.413(1, 4, -22)&0.574& 100.745(2, 1, -71)&0.834&  62.928(3, 0,-163)&0.917&  99.587(3, 1,-102)&0.918& 222.101(3, 4, -43)&0.921\\
 188.195(1, 4, -21)&0.508& 101.779(2, 1, -70)&0.834&  63.327(3, 0,-162)&0.917& 100.590(3, 1,-101)&0.918& 226.755(3, 5, -43)&0.928\\
 197.617(1, 4, -20)&0.511& 103.176(2, 1, -69)&0.833&  63.740(3, 0,-161)&0.917& 101.462(3, 1,-100)&0.919& 230.385(3, 5, -42)&0.929\\
 206.042(1, 4, -19)&0.807& 104.775(2, 1, -68)&0.834&  64.147(3, 0,-160)&0.917& 102.291(3, 1, -99)&0.918& 234.691(3, 5, -41)&0.933\\
 209.377(1, 4, -18)&0.654& 106.364(2, 1, -67)&0.835&  64.526(3, 0,-159)&0.917& 103.302(3, 1, -98)&0.917& 236.404(3, 5, -40)&0.941\\
 219.395(1, 5, -18)&0.522& 107.611(2, 1, -66)&0.841&  64.889(3, 0,-158)&0.917& 104.414(3, 1, -97)&0.916& 241.499(3, 5, -39)&0.920\\
 229.253(1, 5, -17)&0.544& 108.852(2, 1, -65)&0.839&  65.284(3, 0,-157)&0.917& 105.515(3, 1, -96)&0.915& 247.844(3, 5, -38)&0.920\\
 238.302(1, 5, -16)&0.793& 110.550(2, 1, -64)&0.837&  65.716(3, 0,-156)&0.917& 106.425(3, 1, -95)&0.913& 254.230(3, 5, -37)&0.923\\
 243.548(1, 6, -16)&0.654& 112.381(2, 1, -63)&0.839&  66.161(3, 0,-155)&0.917& 107.278(3, 1, -94)&0.914& 259.605(3, 6, -37)&0.929\\
 256.815(1, 6, -15)&0.519& 114.043(2, 1, -62)&0.853&  66.596(3, 0,-154)&0.917& 108.387(3, 1, -93)&0.915& 264.803(3, 6, -36)&0.926\\
 269.995(1, 6, -14)&0.884& 115.304(2, 1, -61)&0.856&  66.998(3, 0,-153)&0.917& 109.553(3, 1, -92)&0.916& 267.478(3, 6, -35)&0.957\\
                   &     & 117.013(2, 1, -60)&0.842&  67.389(3, 0,-152)&0.917& 110.504(3, 1, -91)&0.918& 272.259(3, 6, -34)&0.921\\
  60.415(2, 0,-120)&0.833& 119.068(2, 1, -59)&0.839&  67.822(3, 0,-151)&0.917& 111.412(3, 1, -90)&0.919&\\
\hline
\end{tabular}
\end{table*}

\begin{table*}
\footnotesize
\caption{\label{t7}Comparsions of results of the multiplets in Table 2. $\nu^{\rm obs}$ denotes the observed frequencies in $\mu$Hz, $\nu^{\rm theo}$ denotes the calculated frequencies in $\mu$Hz. $\Delta\nu$ = $|\nu^{\rm obs}-\nu^{\rm theo}|$}
\centering
\begin{tabular}{lccccccccccc}
\hline\hline

Multiplet&ID     &$\nu^{\rm obs}$&$\nu^{\rm theo}$ &$\Delta\nu$   &Multiplet &ID   &$\nu^{\rm obs}$&$\nu^{\rm theo}$&$\Delta\nu$\\
         &       &($\mu$Hz)  &($\mu$Hz)       &($\mu$Hz)     &          &     &($\mu$Hz)  &($\mu$Hz)      &($\mu$Hz)\\
\hline
        &$f_{10}$&106.152   &106.024$(1,-1)$  &0.128         &      &$f_{18}$ &122.559   &123.228$(2,-2)$ &0.669\\
        &        &          &                 &              &11    &         &          &\\
1       &$f_{12}$&110.637   &110.476$(1,0)$   &0.161         &      &$f_{28}$ &144.934   &144.904$(2,+1)$ &0.030\\
        &        &          &                 &              &      &\\
        &$f_{14}$&115.036   &114.928$(1,+1)$  &0.108         &      &\\
        &        &          &                 &              &      &$f_{42}$ &194.179   &194.638$(2,-2)$ &0.459\\
        &        &          &                 &              &12    &         &          &\\
        &$f_{31}$&162.625   &162.360$(1,-1)$  &0.265         &      &$f_{47}$ &216.758   &216.547$(2,+1)$ &0.211\\
        &        &          &                 &              &\\
2       &$f_{34}$&167.007   &166.873$(1,0)$   &0.134         &\\
        &        &          &                 &              &      &$f_{48}$ &222.367   &222.005$(2,-2)$ &0.362\\
        &$f_{35}$&171.485   &171.386$(1,+1)$  &0.099         &13    &         &          &\\
        &        &          &                 &              &      &$f_{51}$ &252.079   &251.871$(2,+2)$ &0.208\\
        &        &          &                 &              &\\
        &$f_{41}$&192.909   &193.216$(1,-1)$  &0.307         &\\
        &        &          &                 &              &      &$f_{22}$ &125.296   &125.696$(3,-1)$ &0.400\\
3       &$f_{43}$&197.503   &197.617$(1,0)$   &0.114         &      &         &          &\\
        &        &          &                 &              &14    &$f_{24}$ &133.453   &133.602$(3,0)$  &0.149\\
        &$f_{44}$&201.898   &202.018$(1,+1)$  &0.120         &      &         &          &\\
        &        &          &                 &              &      &$f_{27}$ &141.765   &141.508$(3,+1)$ &0.257\\
        &        &          &                 &              &      &\\
        &$f_{5}$ &87.275    &86.767$(1,-1)$   &0.508         &      &\\
4       &        &          &                 &              &      &$f_{15}$ &115.706   &116.152$(3,-2)$ &0.446\\
        &$f_{6}$ &96.149    &95.707$(1,+1)$   &0.442         &15    &\\
        &        &          &                 &              &      &$f_{20}$ &123.812   &124.093$(3,-1)$ &0.281\\
        &        &          &                 &              &      &\\
        &$f_{8}$ &100.779   &100.560$(2,-2)$  &0.219         &      &\\
        &        &          &                 &              &      &$f_{17}$ &117.666   &118.118$(3,-3)$ &0.452\\
5       &$f_{11}$&108.372   &107.932$(2,-1)$  &0.440         &16    &\\
        &        &          &                 &              &      &$f_{25}$ &134.071   &133.964$(3,-1)$ &0.107\\
        &$f_{16}$&115.872   &115.304$(2,0)$   &0.568         &      &\\
        &        &          &                 &              &      &\\
        &        &          &                 &              &      &$f_{49}$ &233.083   &233.576$(3,-1)$ &0.493\\
        &$f_{9}$ &102.072   &102.510$(2,-2)$  &0.438         &17    &\\
        &        &          &                 &              &      &$f_{50}$ &249.725   &249.422$(3,+1)$ &0.303\\
6       &$f_{21}$&124.571   &124.264$(2,+1)$  &0.307         &      &\\
        &        &          &                 &              &      &\\
        &$f_{23}$&132.028   &131.516$(2,+2)$  &0.512         &      &$f_{26}$ &134.762   &135.456$(3,0)$  &0.694\\
        &        &          &                 &              &18    &\\
        &        &          &                 &              &      &$f_{30}$ &158.977   &159.148$(3,+3)$ &0.171\\
        &$f_{2}$ &65.541    &65.633$(2,-2)$   &0.092         &      &\\
7       &        &          &                 &              &      &\\
        &$f_{3}$ &72.978    &72.807$(2,-1)$   &0.171         &      &$f_{1}$  &64.936    &65.120$(3,-1)$  &0.184\\
        &        &          &                 &              &19    &\\
        &        &          &                 &              &      &$f_{7}$  &96.938    &96.675$(3,+3)$  &0.263\\
        &$f_{32}$&164.262   &164.281$(2,-2)$  &0.019         &      &\\
8       &        &          &                 &              &      &\\
        &$f_{36}$&171.638   &171.635$(2,-1)$  &0.003         &      &$f_{19}$ &122.769   &122.827$(3,-1)$ &0.058\\
        &        &          &                 &              &20    &\\
        &        &          &                 &              &      &$f_{29}$ &155.380   &154.588$(3,+3)$ &0.792\\
        &$f_{33}$&164.855   &164.831$(2,0)$   &0.024         &      &\\
9       &        &          &                 &              &      &\\
        &$f_{37}$&172.243   &172.143$(2,+1)$  &0.100         &      &$f_{38}$ &176.285   &177.108$(3,-2)$ &0.823\\
        &        &          &                 &              &21    &\\
        &        &          &                 &              &      &$f_{46}$ &209.708   &208.766$(3,+2)$ &0.942\\
        &$f_{39}$&189.056   &188.534$(2,0)$   &0.522         &      &\\
10      &        &          &                 &              &      &\\
        &$f_{45}$&203.652   &203.106$(2,+2)$  &0.546         &      &\\
\hline
\end{tabular}
\end{table*}

\begin{table*}
\caption{\label{t8}Possible mode identifications for the unidentified observed frequencies based on the optimal model. $\Delta\nu$ = $|\nu^{\rm obs}-\nu^{\rm theo}|$.}
\centering
\begin{tabular}{llll}
\hline\hline
ID &$\nu^{\rm obs}$&$\nu^{\rm theo}(l,n_{\rm p},n_{\rm g},m)$ &$\Delta\nu$\\
        &($\mu$Hz)&($\mu$Hz)           &($\mu$Hz)\\
\hline
$f_{13}$&112.291  &112.212(0,1,0)      &0.079\\
        &\\
$f_{4}$ &76.363   &76.394(1,0,-51,-1)  &0.031\\
        &         &76.303(2,0,-117,+2) &0.060\\
        &\\
$f_{40}$&190.612  &190.803(2,3,-39,+2) &0.191\\
        &\\
$f_{52}$&262.964  &262.382(1,6,-14,-1) &0.582\\
\hline
\end{tabular}
\end{table*}

\end{document}